# Artificial neural network for myelin water imaging

Jieun Lee[1]  |  Doohee Lee[1]  |  Joon Yul Choi[1,2]  |  Dongmyung Shin[1]  |  Hyeong-Geol Shin[1]  |  Jongho Lee[1]

[1]Laboratory for Imaging Science and Technology, Department of Electrical and Computer Engineering, Seoul National University, Seoul, Republic of Korea

[2]Cleveland Clinic, Epilepsy Center, Neurological Institute, OH, USA

**Correspondence**

Jongho Lee, Ph.D., Department of Electrical and Computer Engineering, Seoul National University, Building 301, Room 1008, 1 Gwanak-ro, Gwanak-gu, Seoul, Republic Korea.

Email: jonghoyi@snu.ac.kr

**Purpose**: To demonstrate the application of artificial-neural-network (ANN) for real-time processing of myelin water imaging (MWI).

**Methods**: Three neural networks, ANN-$I_{MWF}$, ANN-$I_{GMT2}$, and ANN-II, were developed to generate MWI. ANN-$I_{MWF}$ and ANN-$I_{GMT2}$ were designed to output myelin water fraction (MWF) and geometric mean $T_2$ ($GMT_{2,IEW}$), respectively whereas ANN-II generates a $T_2$ distribution. For the networks, gradient and spin echo data from 18 healthy controls (HC) and 26 multiple sclerosis patients (MS) were utilized. Among them, 10 HC and 12 MS had the same scan parameters and were used for training (6 HC and 6 MS), validation (1 HC and 1 MS), and test sets (3 HC and 5 MS). The remaining data had different scan parameters and were applied to exam the effects of the scan parameters. The network results were compared with those of conventional MWI in the white matter mask and regions of interest (ROI).

**Results**: The networks produced highly accurate results, showing averaged normalized root-mean-squared error under 3% for MWF and 0.4% for $GMT_{2,IEW}$ in the white matter mask of the test set. In the ROI analysis, the differences between ANNs and conventional MWI were less than 0.1% in MWF and 0.1 ms in $GMT_{2,IEW}$ (no statistical difference and $R^2 > 0.97$). Datasets with different scan parameters showed increased errors. The average processing time was 0.68 sec in ANNs, gaining 11,702 times acceleration in the computational speed (conventional MWI: 7,958 sec).

**Conclusion**: The proposed neural networks demonstrate the feasibility of real-time processing for MWI with high accuracy.

**K E Y W O R D S**

Myelin water imaging, artificial neural network, $T_2$ distribution, multi-echo spin echo, multiple sclerosis, deep learning

## 1 | INTRODUCTION

Myelin water imaging (MWI) is an MRI technique that acquires a signal from water confined in the gap between myelin lipid bilayers[1]. This signal, which is referred to as the myelin water signal, has distinguishably shorter $T_2$ relaxation than those of axonal and extra-cellular water signals. As a result, one can generate a voxel-wise $T_2$ distribution of the water compartments by measuring $T_2$ decay[1]. From this $T_2$ distribution, quantitative metrics such as myelin water fraction, which is the ratio of myelin water signal to total water signal, and geometric mean $T_2$ ($GMT_{2,IEW}$), which is the geometric mean of long $T_2$ signal, are extracted to explore the integrity of white matter[2-6]. Unfortunately, data processing to generate the $T_2$ distribution is computationally expensive when correcting for stimulated echoes[7]. It often takes several hours for the processing of the whole brain data. As a result, this data processing is performed off-line, hampering the ability to ensure the quality of myelin water images during a scan session.

In recent years, artificial neural networks (ANN) have been proposed as a promising tool to process complex biomedical data[8]. ANN is known to provide a good approximation for complex functions and is computationally efficient[9]. Additionally, computation in ANN is commonly performed using a graphical processing unit (GPU), which massively parallelizes computation to boost efficiency. By taking these advantages, ANN has been applied for a number of data processing tasks including curve fitting[10,11] and inverse problems[12-15] and has demonstrated the ability to process large size data in substantially shorter processing time than that of conventional methods.

In this study, we took advantage of the computational efficiency of ANN to demonstrate the feasibility of generating whole-brain MWI in less than a second. Three different networks, ANN-I$_{MWF}$, ANN-I$_{GMT2}$, and ANN-II, were developed. ANN-I was designed to generate MWF (ANN-I$_{MWF}$) and $GMT_{2,IEW}$ (ANN-I$_{GMT2}$) maps directly from $T_2$ decay data. On the other hand, ANN-II generated a voxel-wise $T_2$ distribution from which MWF and $GMT_{2,IEW}$ values were calculated.

Source codes for our ANNs are available at https://github.com/snu-list/ANN-MWI.

## 2 | METHODS

### 2.1 | MRI data

MRI data from previously published studies[6,16] were used. The data were from 18 healthy controls (HC) (7 males and 11 females; mean age = 35.7 ± 7.2 years) and 26 multiple sclerosis (MS) patients (11 males and 15 females; mean age = 34.2 ± 6.5 years). The subjects were scanned at a 3T Trio MRI scanner (Siemens Healthcare, Erlangen, Germany) using a 32-channel phased-array head coil under the approval of the institutional review board.

For MWI, a 3D multiple echo gradient and spin echo (GRASE) sequence, which was proposed for MWI[17], was utilized. The scan parameters were as follows: FOV = 240 × 180 × 112 mm$^3$, voxel dimensions = 1.5 × 1.5 × 4 mm$^3$, number of slices = 28, TR = 1000 ms, number of echoes = 32, echo-planar imaging factor = 3, flip angle = 90°, and acquisition time = 14 min 5 sec. The default scan parameters for the first echo time (TE$_1$) and echo spacing were set to be 10 ms each and were used in 22 subjects (10 HC and 12 MS). For the other subjects (remaining 22 subjects), slightly longer TEs were used (TE$_1$ = 10.1 ms for 3 HC and 8 MS, and TE$_1$ = 10.2 ms for 5 HC and 6 MS). To meet a specific absorption rate limit, longer TRs were used for a few subjects (see Supporting Information Table S1).

Additionally, three clinical protocol scans were used to detect lesions: 2D $T_1$-weighted spin echo imaging (in-plane resolution = 0.8 × 0.7 mm$^2$; slice thickness = 3.2 mm; number of slices = 32; TR = 550 ms; TE = 9.2 ms; and flip angle = 70°), 2D $T_2$-weighted fast spin echo imaging (in-plane resolution = 0.5 × 0.5 mm$^2$; slice thickness = 3.2 mm; number of slices = 32; TR = 8750 ms; echo train length = 21; TE = 90 ms; and echo spacing = 11.3 ms), and 2D FLAIR imaging (in-plane resolution = 0.7 × 0.7 mm$^2$; slice thickness = 3.2 mm; number of slices = 32; TR = 9000 ms; TE = 87 ms; TI = 2500 ms; echo train length = 16; and echo spacing = 8.74 ms).

### 2.2. Conventional MWI

Using the multi-echo GRASE data, MWI was generated as a reference for ANNs. The data processing started with a Tukey window (coefficient = 0.33) applied to the k-space of the multi-echo images in order to suppress Gibb's artifacts. Then voxel-wise multi-echo data were processed to generate a $T_2$ distribution by fitting stimulated-echo corrected multi-exponential functions[7]. The following parameters, which were common for clinical studies[18-21], were used for the fitting: number of exponential functions = 120; $T_2$ range = 15 to 2000 ms, logarithmically spaced; and chi-square regularization with the constraint of $1.020\chi^2_{min} \leq \chi^2 \leq 1.025\chi^2_{min}$. From the $T_2$ distribution, MWF was calculated by dividing the sum of the signals from 15 to 40 ms by the sum of the entire $T_2$ distribution. Additionally, the geometric mean $T_2$ of the main water peak between 40 ms and 200 ms was calculated using

$$\text{GMT}_{2,\text{IEW}} = \exp\left[\sum_{j=M_1}^{M_2} S(T_{2,j}) \log T_{2,j} \Big/ \sum_{j=M_1}^{M_2} S(T_{2,j})\right] \quad (1)$$

where $\text{GMT}_{2,\text{IEW}}$ is the geometric mean $T_2$, $S(T_{2,j})$ is the amplitude of the $T_2$ distribution at $T_{2,j}$, and $j = M_1$ and $j = M_2$ correspond to $T_2$ of 40 ms and 200 ms, respectively[22]. This approach of generating MWI is referred to as conventional MWI hereafter.

### 2.3. Artificial neural networks for MWI

In this work, three different artificial neural networks, ANN-I$_{MWF}$, ANN-I$_{GMT2}$, and ANN-II, were developed (Fig. 1).

ANN-I was designed to generate MWF (ANN-I$_{MWF}$) or GMT$_{2,\text{IEW}}$ (ANN-I$_{GMT2}$) directly from $T_2$ decay data by training the network with the 32-echo GRASE data of a voxel as an input and the MWF or GMT$_{2,\text{IEW}}$ from the conventional MWI as a label (Fig. 1b). The networks for MWF and GMT$_{2,\text{IEW}}$ were trained separately, generating ANN-I$_{MWF}$ and ANN-I$_{GMT2}$. The networks had 32 neurons in the input layer and one neuron in the output layer. Between the input and output layers, seven hidden layers were constructed with 160, 240, 320, 360, 480, 520, and 600 fully connected neurons. A leaky rectified linear unit (alpha = 0.2) was used as a non-linear activation function[23]. We utilized Adam optimizer[24] with a varying learning rate, which started from 0.001 and reduced by one-tenth at every predetermined epoch (900, 1200, 1500, and 1800 epochs) in order to enhance the training speed with more accurate generalization[25]. The batch size was increased by one for each epoch from 2 to 2,002. This approach has shown to decrease the number of parameter updates and improve generalization and training performance[26]. To avoid potential overfitting, the early-stopping method[27], a common approach to prevent overfitting, was used. The loss function was defined as the mean squared error between the network output and the label data.

ANN-II was designed to generate a $T_2$ distribution by training the network with the 32-echo data of a voxel as an input and the $T_2$ distribution from the conventional MWI as a label (Fig. 1c). The network had the same structure as ANN-I except for the output layer, which had 120 neurons to represent the coefficients of the 120 exponential basis functions. The loss function was defined as the mean squared error between the network output and the $T_2$ distribution from the conventional MWI. The other hyper-parameters were the same as ANN-I. To enforce the non-negative nature of the $T_2$ distribution, the negative

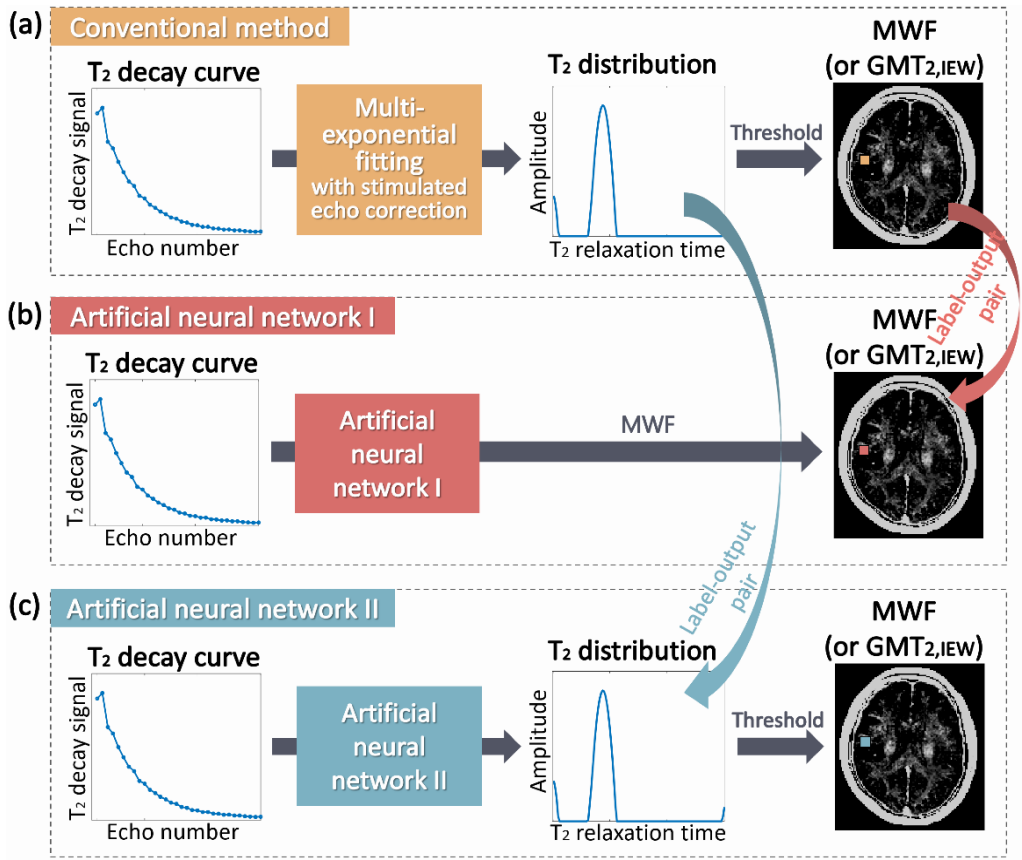

**Figure 1.** Conceptual illustration of the conventional MWI method (a), artificial neural network 1 (ANN-I$_{MWF}$ and ANN-I$_{GMT2}$; b), and artificial neural network 2 (ANN-II; c). ANN-I was trained to generate MWF (ANN-I$_{MWF}$) or GMT$_{2,IEW}$ (ANN-I$_{GMT2}$) from the T$_2$ decay curve whereas ANN-II was trained to generate the T$_2$ distribution. The T$_2$ decay curve shows a characteristic decay with the second echo signal larger than the first echo signal due to the stimulated echo in the presence of B$_1$ inhomogeneity.

values in the output were forced to zero.

For all networks, the input and output training data were normalized to improve training accuracy and learning speed[28]. For the input, the 32-echo data were divided by the first echo. In ANN-I$_{MWF}$ and ANN-I$_{GMT2}$, the GMT$_{2,IEW}$ values were divided by 100 while keeping the MWF values the same. For the output of ANN-II, the T$_2$ distribution was scaled to have the sum of the T$_2$ distribution to be 15, which was chosen empirically to improve training accuracy.

During network performance optimization, various data combinations were tested using the 22 subject datasets of the default scan parameters (10 HC and 12 MS; see Supporting Information). Finally, out of the 22 datasets, twelve datasets (6 HC and 6 MS) were used as a network training set, eight datasets (3 HC and 5 MS) were reserved as a test set, and two datasets (1 HC and 1 MS) were utilized as a validation set. The remaining datasets of the different scan parameters (22 subjects; TE$_1$ = 10.1 ms for 3 HC and 8 MS; and TE$_1$ = 10.2 ms for 5 HC and 6 MS) were utilized to test the effects of the scan parameters on network performance.

When training and testing the networks, two different masks, a brain mask and a white matter mask, were applied. For the network training, the brain mask, which included gray and white matters, was used. It was created from the FLAIR image by extracting the brain[29] and transforming the result into an MWF map space[30]. For the test, the white matter mask was generated using the T$_1$-weighted image over the T$_2$- weighted image as described in the work of Choi et al.[6]. The mask was refined to exclude voxels with unrealistic MWF (MWF ≥ 30% or MWF = 0). For the MS patients, MS lesion ROI was generated by applying a threshold to the FLAIR image[6] and superimposing it onto the white matter mask. This white matter mask was utilized to evaluate an MWF map. The normalized root-mean-square error (NRMSE) was calculated in the white matter mask for the eight test subjects (both HC and MS) with the conventional MWI as a reference. Both MWF and $GMT_{2,IEW}$ were compared. The total number of voxels for the network training and testing were approximately 1,400,000 and 430,000, respectively.

For quantitative comparison, the analysis was performed for the white matter mask and five regions of interest (ROIs): forceps minor, genu of the corpus callosum, posterior limb of the internal capsule (PLIC), splenium of the corpus callosum, and forceps major. The five ROIs were manually segmented guided by a reference[31]. The three HCs in the test set were used for this analysis. A Wilcoxon signed-rank test was performed between the conventional MWF (or $GMT_{2,IEW}$) and ANN MWF (or $GMT_{2,IEW}$) maps for each ROI. Additionally, the voxel-wise correlation was calculated between the conventional MWF (or $GMT_{2,IEW}$) and ANN MWF (or $GMT_{2,IEW}$) for each ROI. Finally, a Bland-Altman plot was plotted for the five ROIs to analyze the agreement of MWF (or $GMT_{2,IEW}$).

For MS patients, MWF and $GMT_{2,IEW}$ were compared in the lesions of the five MS test set. Then, a Wilcoxon signed-rank test and the voxel-wise correlation between the conventional MWI and ANNs were performed for the lesion ROI. Additionally, the NRMSE was calculated in the MS lesion ROI.

ANN-II generates a voxel-wise $T_2$ distribution and, therefore, is flexible in choosing a threshold for myelin water. To test the reliability of ANN-II for different thresholds, MWF maps were generated with three different thresholds (30, 40, and 50 ms). The NRMSE was calculated for each threshold using the eight test set.

The network training and test were performed on a GPU workstation (NVIDIA GeForce GTX 1080 Ti GPU with Intel® Xeon(R) CPU E5-2603 v4 at 1.70 GHz) using TensorFlow[32]. The conventional MWI was processed using the four CPU cores of the same workstation and MATLAB R2017b (Mathworks, Inc., Natick, MA). The processing time for the test (or inference) was compared for all methods by repeating the test ten times and averaging the processing time.

## 3 | RESULTS

The MWF and $GMT_{2,IEW}$ maps from the conventional MWI, ANN-I, and ANN-II are shown in Figure 2 for representative HC and MS. Additionally, the error maps, which are the absolute difference between the results of the conventional MWI and ANN MWI multiplied by 10, are displayed. The MWF and $GMT_{2,IEW}$ maps illustrate no visible difference among the three methods. This is also confirmed by the error maps, which reveal noticeable errors only outside of white matter. When the NRMSE is calculated in the white matter mask of the eight test set (both HC and MS), the average errors of MWF are 2.86 ± 0.29% in ANN-$I_{MWF}$ (HC: 3.01 ± 0.03%, MS: 2.77 ± 0.35%), and 2.26 ± 0.20% in ANN-II (HC: 2.33 ± 0.06%, MS: 2.21 ± 0.24%). The average NRMSEs of $GMT_{2,IEW}$ are 0.34 ± 0.03% in ANN-$I_{GMT2}$ (HC: 0.34 ± 0.04%, MS: 0.34 ± 0.04%), and 0.22 ± 0.05% in ANN-II (HC: 0.18 ± 0.02%, MS: 0.24 ± 0.06%).

The results of the ROI analysis in the three HC test set are summarized in Table 1. The mean MWF and $GMT_{2,IEW}$ in the six ROIs, the white matter mask and five ROIs, reveal almost identical results in all methods. No statistically significant difference is observed in MWF and $GMT_{2,IEW}$. Additionally, extremely high correlation values ($R^2$ > 0.99) are reported for the voxel-wise correlation between the results of the ANNs and the conventional method in all ROIs. The Bland-Altman plots are included in Supporting Information Figures S1 and S2, consolidating the results. A few examples of $T_2$ distributions from the conventional MWI and ANN-II are displayed in Supporting Information Figure S3. Each plot corresponds to a single pixel in the five ROIs (Supporting Information Figure S4). The results reconfirm the robust performance of ANN-II in generating voxel-wise $T_2$ distributions.

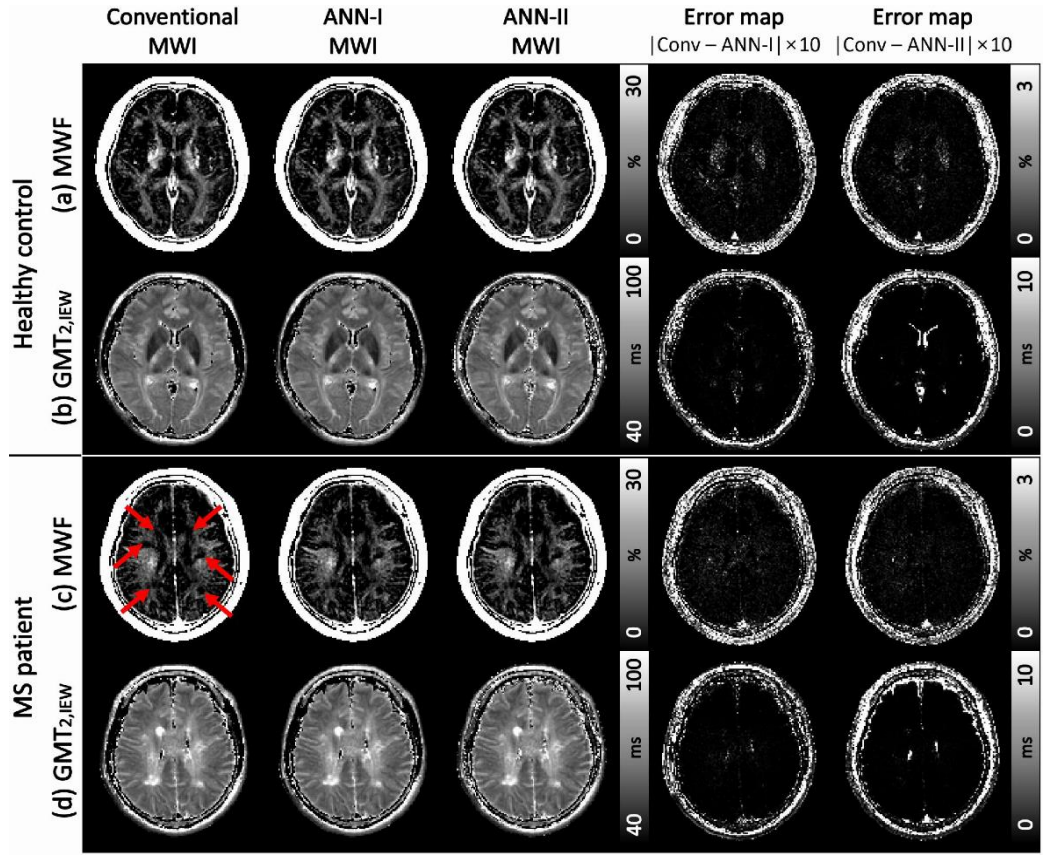

**Figure 2.** Myelin water fraction maps (a and c) and geometric mean $T_2$ maps (b and d) in a healthy control (upper rows) and an MS patient (lower rows) using the conventional MWI, ANN-I, and ANN-II. Error maps, which are multiplied by 10, are displayed. The red arrows indicate an MS lesion.

**Table 1.** ROI analysis results in healthy controls. Mean ± standard deviation of MWF (%) and $GMT_{2,IEW}$ (ms) are reported in the white matter mask and five ROIs. When comparing the results of the conventional MWI and ANN-I or ANN-II, strong voxel-wise correlations ($R^2 > 0.99$) are reported in all ROIs. PLIC represents the posterior limb of the internal capsule.

| Healthy controls (n = 3) | Whole white matter | Forceps minor | Forceps major | Genu | Splenium | PLIC |
|---|---|---|---|---|---|---|
| **(a) Conventional MWF (%)** | 7.4 ± 0.3 | 7.5 ± 0.9 | 11.5 ± 2.3 | 12.3 ± 2.7 | 15.2 ± 1.5 | 18.3 ± 1.3 |
| **(b) ANN-I MWF (%)** | 7.4 ± 0.3 | 7.5 ± 0.9 | 11.5 ± 2.3 | 12.2 ± 2.8 | 15.2 ± 1.4 | 18.3 ± 1.3 |
| **(c) ANN-II MWF (%)** | 7.4 ± 0.3 | 7.5 ± 0.9 | 11.5 ± 2.3 | 12.3 ± 2.7 | 15.2 ± 1.5 | 18.3 ± 1.3 |
| $R^2$ correlation bewteen (a) and (b) | 0.99 | 0.99 | 0.99 | 0.99 | 0.99 | 0.99 |
| $R^2$ correlation between (a) and (c) | 0.99 | 0.99 | 0.99 | 0.99 | 0.99 | 0.99 |
| **(d) Conventional $GMT_{2,IEW}$ (ms)** | 63.2 ± 1.3 | 59.9 ± 2.3 | 72.6 ± 4.0 | 58.3 ± 3.8 | 67.3 ± 3.9 | 72.7 ± 4.5 |
| **(e) ANN-I $GMT_{2,IEW}$ (ms)** | 63.2 ± 1.3 | 59.9 ± 2.3 | 72.6 ± 4.0 | 58.3 ± 3.8 | 67.3 ± 3.9 | 72.7 ± 4.4 |
| **(f) ANN-II $GMT_{2,IEW}$ (ms)** | 63.2 ± 1.3 | 59.9 ± 2.3 | 72.6 ± 4.0 | 58.3 ± 3.8 | 67.3 ± 3.9 | 72.7 ± 4.5 |
| $R^2$ correlation between (d) and (e) | 1.00 | 0.99 | 0.99 | 0.99 | 0.99 | 0.99 |
| $R^2$ correlation between (d) and (f) | 1.00 | 1.00 | 1.00 | 1.00 | 1.00 | 0.99 |

**Table 2.** ROI analysis results in MS patients. Mean ± standard deviation of MWF (%) and $GMT_{2,IEW}$ (ms) are reported in the MS lesions. When comparing the results of the conventional MWI and ANN-I or ANN-II, strong voxel-wise correlations ($R^2$ > 0.97) are reported in all ROIs.

| MS patients (n = 5) | Patient 1 lesion | Patient 2 lesion | Patient 3 lesion | Patient 4 lesion | Patient 5 lesion |
|---|---|---|---|---|---|
| (a) Conventional MWF (%) | 6.7 ± 4.3 | 6.8 ± 5.2 | 3.9 ± 2.6 | 5.2 ± 3.3 | 3.1 ± 4.3 |
| (b) ANN-I MWF (%) | 6.7 ± 4.3 | 6.8 ± 5.2 | 3.9 ± 2.6 | 5.2 ± 3.3 | 3.1 ± 4.3 |
| (c) ANN-II MWF (%) | 6.6 ± 4.3 | 6.8 ± 5.2 | 3.9 ± 2.6 | 5.2 ± 3.3 | 3.1 ± 4.3 |
| $R^2$ correlation between (a) and (b) | 0.99 | 0.99 | 0.98 | 0.99 | 0.99 |
| $R^2$ correlation between (a) and (c) | 0.99 | 0.99 | 0.98 | 0.99 | 0.99 |
| (d) Conventional $GMT_{2,IEW}$ (ms) | 98.7 ± 19.6 | 84.2 ± 13.8 | 96.8 ± 15.5 | 91.0 ± 16.9 | 91.8 ± 12.6 |
| (e) ANN-I $GMT_{2,IEW}$ (ms) | 98.6 ± 19.5 | 84.2 ± 13.8 | 96.7 ± 15.5 | 91.1 ± 16.9 | 91.7 ± 12.6 |
| (f) ANN-II $GMT_{2,IEW}$ (ms) | 98.7 ± 19.1 | 84.2 ± 13.8 | 96.7 ± 15.5 | 91.0 ± 16.8 | 91.7 ± 12.5 |
| $R^2$ correlation between (d) and (e) | 0.99 | 0.99 | 0.99 | 0.99 | 0.99 |
| $R^2$ correlation between (d) and (f) | 0.97 | 0.99 | 0.99 | 0.99 | 0.99 |

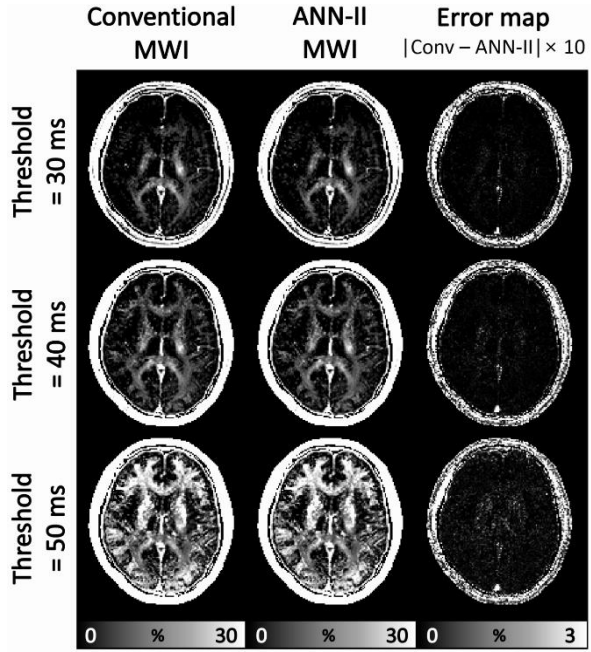

**Figure 3.** MWF maps from the three different thresholds (30, 40, and 50 ms) with error maps multiplied by 10. The MWF maps show similar contrasts in all cases, demonstrating the robustness of ANN-II in generating the $T_2$ distribution.

When the analysis is performed for the MS lesion ROI in the five MS test set, the same trends are observed in all metrics (Table 2). The mean NRMSEs of MWF in the lesion ROI are 4.92 ± 0.88% in ANN-$I_{MWF}$, and 4.59 ± 0.87% in ANN-II, whereas those of $GMT_{2,IEW}$ are 0.63 ± 0.15% in ANN-$I_{GMT2}$, and 0.58 ± 0.13% in ANN-II. These results are higher than those from the white matter mask (see Discussion).

Figure 3 shows the MWF maps from the three different myelin water thresholds (30, 40, and 50 ms), demonstrating the flexibility and robustness of ANN-II in generating MWF maps with the different thresholds. The mean NRMSEs in the white matter mask of the eight test data are 2.65 ± 0.34%, 2.26 ± 0.20%, and 2.27 ± 0.19% for the thresholds of 30, 40, and 50 ms, respectively.

When the datasets with different TEs are applied to the network, they result in increased errors of MWI when compared to the datasets with $TE_1$ = 10 ms. For the datasets with $TE_1$ = 10.1 ms, the NRMSEs of MWI are 4.32 ± 0.48% in ANN-I$_{MWF}$ (HC: 4.24 ± 0.18%, MS: 4.35 ± 0.56%), and 4.02 ± 0.56% in ANN-II (HC: 4.05 ± 0.13%, MS: 4.01 ± 0.67%). For TE$_1$ = 10.2 ms datasets, the NRMSEs of MWI are 8.02 ± 0.84% in ANN-I$_{MWF}$ (HC: 8.00 ± 0.58%, MS: 8.04 ± 1.07%), and 7.90 ± 0.83% in ANN-II (HC: 7.92 ± 0.55%, MS: 7.88 ± 1.07%).

When the processing times of the whole brain data are compared, the ANN methods (0.68 sec) are approximately 11,702 times faster than the conventional method (7,958 sec or 2.2 hours), demonstrating the feasibility of applying the neural networks for real-time processing of MWI.

## 4 | DISCUSSION

In this study, we developed fast and robust data processing approaches for MWI by using ANN. The results showed under 3% of average NRMSE in MWF and 0.4% in GMT$_{2,IEW}$ while gaining 11,702 times faster computational speed (less than 1 sec for ANNs vs. over 2.2 hours for conventional MWI).

As summarized in Tables 1 and 2, the maximum difference in the ROI-averaged MWF was less than 0.1% (0.06%) when comparing the conventional MWI and ANNs. No statistically significant difference was reported in all ROIs. The same results were observed for the GMT$_{2,IEW}$ results, revealing the maximum difference of less than 0.1 ms (0.09 ms). These results suggest that ANNs may be applied for an ROI analysis with high reliability.

Although the inference of the ANNs can be performed almost in real-time, the network training took approximately 12 hours in our single GPU workstation. This training is required once and, therefore, does not hamper the real-time processing of new data.

In this work, the processing speeds of the ANNs and the conventional MWI were compared using the processors that were optimized for each processing (i.e., one GPU for the ANNs and quad-core CPU for the conventional MWI). When the comparison was performed using the same processor (one CPU core) for all methods, the ANNs took 25.2 sec whereas the conventional MWI took 28,250 sec. In this case, the computational speed of the ANNs was 1121 times faster than that of the conventional MWI, confirming the computational efficiency of the ANNs.

During the development of the networks, optimization was performed for the training of different subject types (Supporting Information Figure S5) and different numbers of subjects (Supporting Information Figure S6). When the effects of the subject type were explored using three different compositions of training sets (6 HC only; 6 MS only; 3 HC and 3 MS combined) for three ROIs (white matter in HC; white matter in MS excluding MS lesions; MS lesions), the networks that included MS patients for training showed less errors in the MS lesions (Supporting Information Figure S5). When the effect of the training data size was investigated by increasing the training data size from 2 to 12 subjects with an equal number of HC and MS, the NRMSE showed the best results at 12 subjects (Supporting Information Figure S6). These two optimization results led us to train the networks using the 12 subjects (6 HC and 6 MS).

To evaluate the robustness of the proposed methods to noise, three different levels of noise (one, two, and three times of noise standard deviation in the GRASE data) were added to the GRASE data (see Supporting Information Table S2). The results revealed that degradations in NRMSE were consistent in both conventional method and ANNs.

The loss function for ANN-II was defined as the mean squared error between the network output and the T$_2$ distribution from the conventional MWI. No weighting was applied. When tested for larger weights in the short T$_2$ components, no significant improvement was observed.

In our results, larger NRMSEs were observed in the MS lesions when compared to those from the white matter mask of the test set. This performance degradation may be explained by an imbalanced number of voxels between the lesion and non-lesion white matter since the MS lesions were approximately 0.3% of the total number of data[33]. Further reduction in the NRMSE may be achieved by balancing training data between non-lesion and lesion by oversampling lesion data[34].

When normalizing the data for the networks, the multi-echo GRASE data of each voxel was divided by the first echo signal in order to set the range of the data approximately from 0 to 1[28]. A logical approach for data normalization is to set the signal at TE of 0 ms being equal to the sum of the $T_2$ distribution. However, the $T_2$ distribution is not available for inference data and, therefore, multi-echo data cannot be normalized using the distribution. As an alternative option, we set the first echo to 1. In ANN-II, the $T_2$ distribution was scaled to have the sum of the $T_2$ distribution to be 15. This scaling does not affect MWF and $GMT_{2,IEW}$ values since they are defined as the ratios of the $T_2$ distribution.

The decay curve in Figure 1 displays a typical pattern in our GRASE MWI data and reveals a larger second echo signal than the first echo signal. This is due to the stimulated echo from imperfect $B_1$ homogeneity. The conventional method demonstrated robustness to $B_1$ inhomogeneity when corrected for the stimulated echo[7]. Since our ANNs are trained using real data with stimulated echoes, we expect similar robustness. However, further investigation is necessary.

When fitting the multi-exponential function, the chi-square regularization of the given parameters (i.e., $1.020\chi^2_{min} \leq \chi^2 \leq 1.025\chi^2_{min}$) is commonly used[35-38]. Changing this factor has an impact on the results of the $T_2$ decomposition[39]. Therefore, the networks need to be retrained if new regularization parameters are desirable.

In our results, the data with different TEs showed increased errors, suggesting the dependency of the ANNs on $TE_1$ and echo spacing. This outcome may be explained by $T_2$ decay variations for the different $TE_1$ and echo spacing. Considering the increased errors for the small increases in TE, a larger difference in TE may introduce substantial degradation in the performance. To reduce the errors, one may train a network for each TE at the cost of increased training datasets. When training different TE datasets, transfer learning can be used to reduce the size of training datasets[40].

The ANNs may be applied for the diagnosis of other diseases such as neuromyelitis optica[20], schizophrenia[42], and phenylketonuria[43]. However, further tests may be necessary to confirm the reliability of the results because errors may increase for untrained lesion types that have different $T_2$ relaxation characteristics. If error increases, one may fine-tune the network with a few datasets using a transfer learning method[40] to improve the performance.

# 5 | CONCLUSIONS

In this work, we proposed three different neural networks for the real-time processing of MWI. The accuracy of the networks in estimating MWF and $GMT_{2,IEW}$ was close to the results of the conventional MWI. The gain in the computational speed was almost 10,000 times. The proposed networks were capable of estimating not only MWF and $GMT_{2,IEW}$ (both ANN-I and ANN-II) but also $T_2$ distribution (ANN-II), and were applicable to healthy controls and MS patients. Our results demonstrated the potentials of applying a neural network for myelin water imaging.

## ACKNOWLEDGMENTS
This work was supported by the National Research Foundation of Korea Grant funded by the Korea government (MSIT) (NRF-2018R1A2B3008445 and

## SUPPORTING INFORMATION

**Table S1.** Summary of TR for each $TE_1$ with the number of subjects.

|  | $TE_1$ (ms) | | |
| --- | --- | --- | --- |
|  | 10 | 10.1 | 10.2 |
| **TR (ms)** | **Number of Subjects** | | |
| **1000 – 1100** | 16 | 8 | 7 |
| **1100 – 1200** | 2 | 2 | 0 |
| **1200 – 1300** | 3 | 0 | 2 |
| **1300 – 1400** | 1 | 0 | 0 |
| **1400 – 1500** | 0 | 1 | 2 |

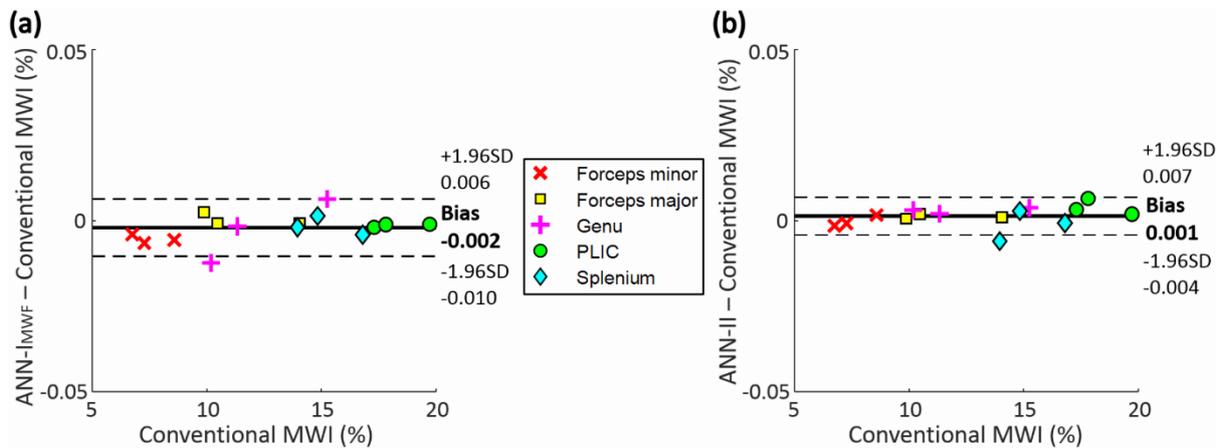

**Figure S1.** Bland-Altman plot comparing MWF values between the conventional MWI and ANN-$I_{MWF}$ (a), and between the conventional MWI and ANN-II (b) in the five ROIs of the three HC test set. The solid line indicates mean difference, and the dashed lines indicate the 95% limits of agreement. The analysis results confirm negligible biases (ANN-$I_{MWF}$: -0.002 and $P = 0.08$; ANN-II: 0.001 and $P = 0.08$).

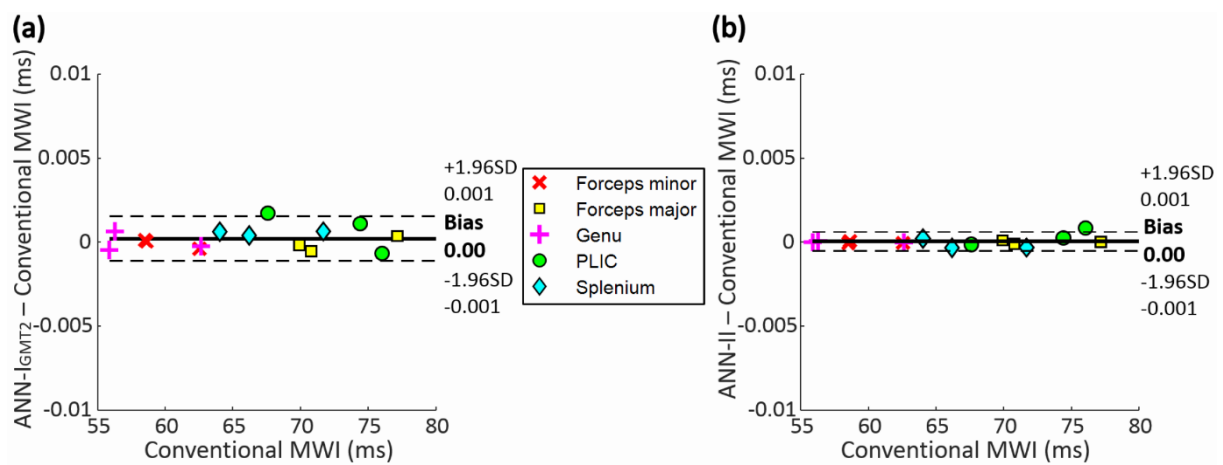

**Figure S2.** Bland-Altman plot comparing $GMT_{2,IEW}$ values between the conventional MWI and ANN-$I_{GMT2}$ (a), and between the conventional MWI and ANN-II (b) in the five ROIs of the three HC test set. The solid line indicates mean difference, and the dashed lines indicate the 95% limits of agreement. The analysis results confirm negligible biases (ANN-$I_{GMT2}$: -0.0002 and $P = 0.24$; ANN-II: 0.00004 and $P = 0.55$).

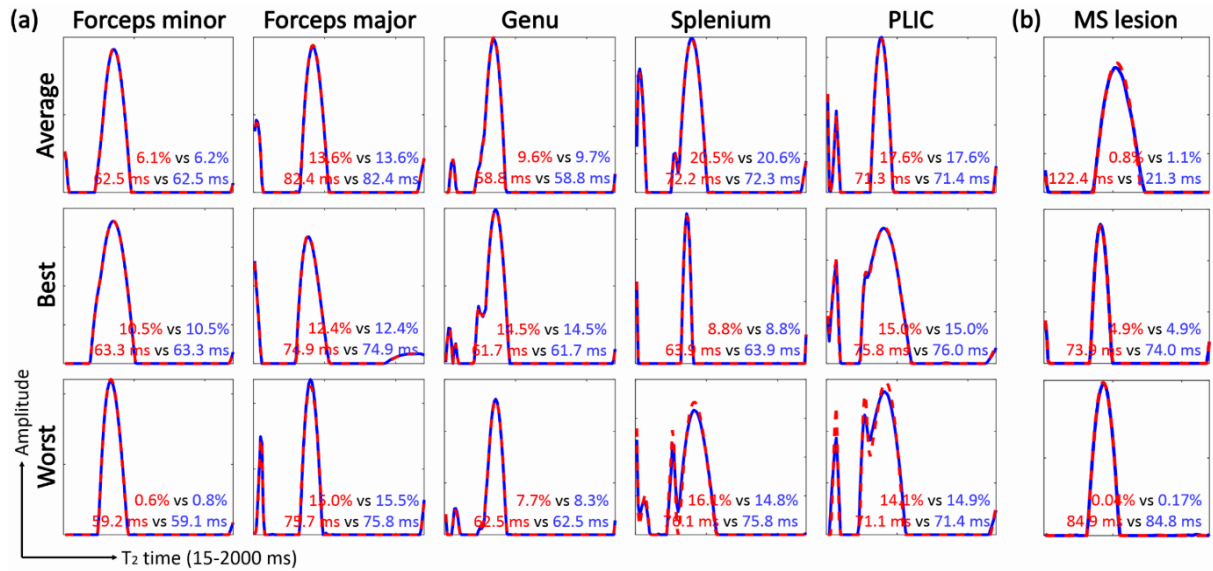

**Figure S3.** $T_2$ distributions of the conventional MWI (dashed red line) and ANN-II (solid blue line) in HC from the five ROIs (a) and MS from an MS lesion (b) in Fig. S4. From each ROI, three pixels that have the best, the worst, and the average NRMSE are selected. The values in each plot report MWF (%) and $GMT_{2,IEW}$ (ms) from the conventional MWI (red) and ANN-II (blue). For comparison, the $T_2$ distributions from the conventional method are scaled in amplitude to match the $T_2$ distribution from ANN-II.

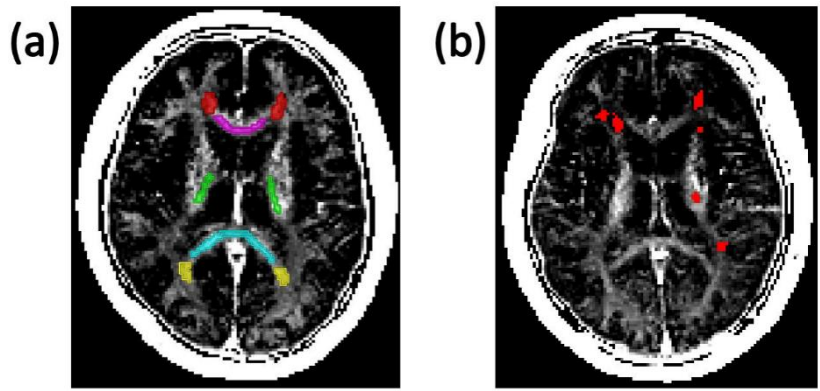

**Figure S4.** Five ROIs, forceps minor (red), genu of the corpus callosum (pink), PLIC (green), splenium of the corpus callosum (blue), and forceps major (yellow) (a), and an MS lesion (red) (b).

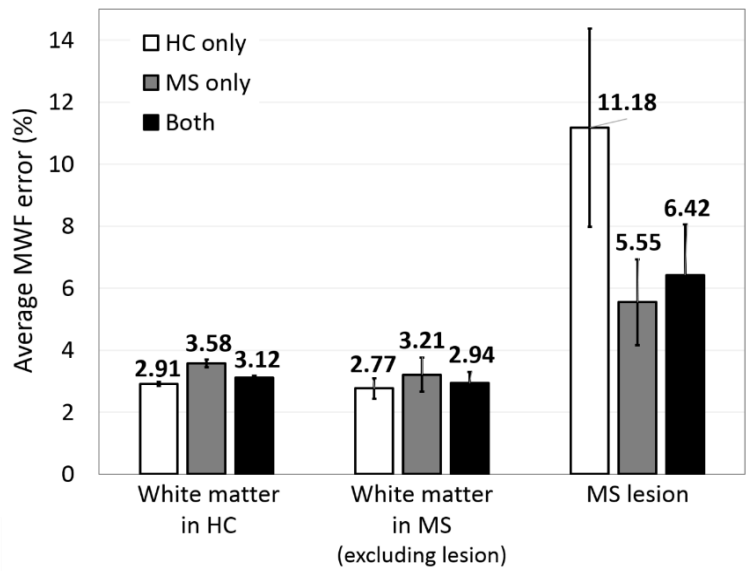

**Figure S5.** Effects of the subject type on network performance. Error bars indicate standard deviation. ANN-II was trained using three different subject compositions: 6 HC only, 6 MS only, and 3 HC and 3 MS combined. Then the performance test, which was measured by the NRMSE of MWF in three ROIs (white matter in HC; white matter in MS excluding lesion; MS lesions), was conducted using the eight test subjects (3 HC and 5 MS). The results reveal that the NRMSE in the MS lesion is higher particularly when the network was trained by HC only.

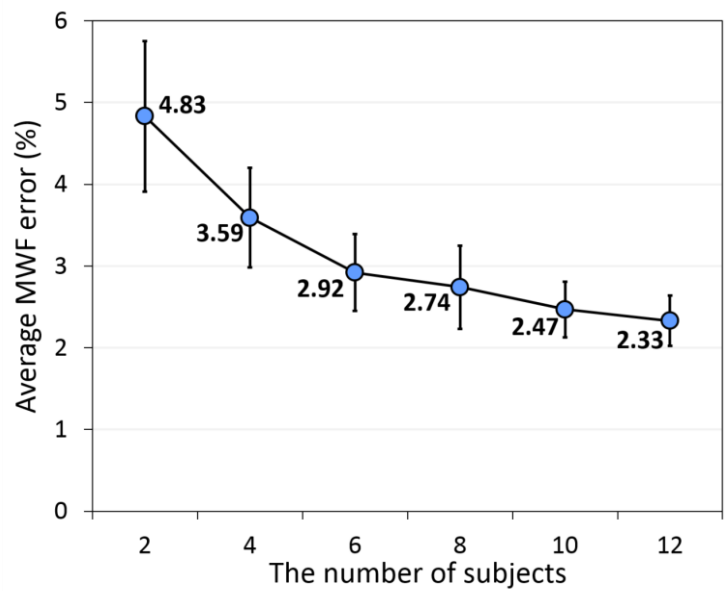

**Figure S6.** Effect of the subject number on network performance. Error bars indicate standard deviation. ANN-II was trained using different numbers of the training set (2 to 12 subjects paired with HC and MS). The performance was measured by the NRMSE of MWF in the whole white matter. K-fold cross-validation was performed[45]. The results from the training set with 12 subjects show the minimum NRMSE.

**Table S2.** Comparison of NRMSE when three different levels of noise were added to the GRASE data of the eight test subjects. The noise-added data were processed using conventional MWI, ANN-I$_{MWF}$, and ANN-II. To demonstrate the degrees of degradation, the mean ± standard deviation of NRMSE is reported in the white matter mask with the original MWF as a reference. The results suggest the noise effects are consistent in all methods.

| NRMSE (%) | Noise SD × 1 added | Noise SD × 2 added | Noise SD × 3 added |
|---|---|---|---|
| **Conventional MWF** | 16.69 ± 2.10% | 29.18 ± 3.26% | 39.29 ± 4.02% |
| **ANN-I MWF** | 16.65 ± 2.09% | 29.09 ± 3.19% | 38.84 ± 3.75% |
| **ANN-II MWF** | 16.64 ± 2.08% | 29.18 ± 3.35% | 39.06 ± 3.93% |